\documentclass[fleqn,10pt]{wlscirep}
\usepackage{longtable}
\usepackage{subfigure}
\usepackage{graphicx}
\usepackage{caption}

\title{Tamper-Evident Complex Genomic Networks}
\author[1,+]{Komal Batool}
\author[2,*,+]{Muaz A. Niazi}

\affil[1]{National University of Science \& Technology, Information Security Department, Islamabad, 44000, Pakistan}
\affil[2]{COMSATS Institute of IT, Computer Science Department, Islamabad, 44000, Pakistan}

\affil[*]{muaz.niazi@ieee.org}

\affil[+]{these authors contributed equally to this work}

\begin{abstract}
Networks are important storage data structures now used to store personal information of individuals around the globe. With the advent of personal genome sequencing, networks are going to be used to store personal genomic sequencing of people. In contrast to social media networks, the importance of relationships in this genomic network is extremely significant. Losing connections between individuals thus implies losing relationship information (E.g. father or son etc.). There currently exists a considerably serious problem in the current approach to storing network data. Simply stated, network data is not tamper-evident. In other words, if some links or nodes were changed/removed/added by a malicious attacker, it would be impossible for the administrator to detect such changes. While, in the current age of social media networks, change in node characteristics and links can be bad in terms of relationships, in the case of networks for storing personal genomes, the results could be truly devastating. Here we present a scheme for building tamper-evident networks using a combination of Cryptographic and Ego-based Network  analytic methods. Using actual published data-sets, we also demonstrate the utility and validity of the scheme besides demonstrating its working in various possible scenarios of usage. Results from the extensive experiments demonstrate the validity of the proposed approach.
\end{abstract}
\begin{document}

\flushbottom
\maketitle
%
%
\thispagestyle{empty}


\section*{Introduction}
In less than a decade, the idea of networks has evolved from being considered as a purely theoretical concept from Computer Science, to being used almost everywhere. Current networks examples include Online Social Networks (OSNs) such as Facebook, Twitter, Reddit, and Google+ among others. The core idea of a Social network (or simply network) is based on connectivity between different entities in the same system \cite{kossinets2006empirical}.  Networks can range from social to biological or computational \cite{altamimi2016towards} in nature \cite{malek2016computational}. Various chaotic effects can be observed in networks \cite{perc2006chaos}.  Often times, it is important to study such complex systems from a multidisciplinary perspective  \cite{trenchard2016equivalences}. Other examples of multidisciplinary approaches include the use of complex systems approach to evaluate emotions in Hollywood \cite{cipresso2016computational} and Brain computer interfaces \cite{ramadan2017brain}. 

The traditional methods of studying complex networks involve developing models and performing analysis, which needs to be validated, as we have previously discussed in \cite{batool2014towards}. The actual effects of tampering in such networks depends primarily on the information contained in the network. Even if the network were merely social network, then unexpected changes in information could result in anything ranging from discontent to severe relationship setbacks. This would be compounded in the cases of individuals connected across multiple networks such as the case of multiplex networks \cite{mucha2010community}.

If however, the data were something as important as personal genomic information of individuals, the implications would be a lot more. This data could be linked with other patients who are  genetically related to each other or else, even relatives unrelated by blood (such as spouses etc.). This is an important change because if there are no mechanism for detecting tampering in a social network, the data in such social network databases would be prone to hacking and tampering besides the need for privacy \cite{gross2005information}. Effects on the virtual perceptions of individuals can also result in severe symptoms in the actual lives of individuals \cite{cipresso2015modeling}.

In the past, various techniques and security mechanisms have been introduced for securing networks. There are a number of ways in which intrusion in networks are detectable based on the features of attacks \cite{demara2004mitigation}, some of which includes detecting the pathway accessed by attacker, exploiting system vulnerabilities and targeting the data or data collection methods in networks \cite{howard1998common}. Previous literature has classified the attacks according to the intents of attackers \cite{bott1999evaluating} as well as based on the level of their knowledge and expertise in penetration in to the systems \cite{anderson1996tamper}. Unfortunately, however social networks still lack mechanisms which give the ability to administrators to verify if anything has been tampered till now. The absence of such complete solutions implies the vulnerability of such networks \cite{leskovec2007cost, memon2006investigative, nath2006crime, o2006using, oatley2006decision, sommer2010outside, van2005process, wang2009framework, yardi2009detecting}. This is especially difficult in genomic social networks because of the large size of the data, often requiring novel techniques to handle it effectively \cite{azar2015dimensionality}. 

The goal of this paper is to present a comprehensive mechanism to ensure tamper-detection in any type of networks, with a specific focus on genomic networks. We also demonstrate the validity of the algorithms by means of an application on the links between social entities of the network. This allows for a focus on the network structure rather than on the attributes of nodes. The contributions of our work can be summarized as follows:

We present a cognitive digital footprinting method allowing for a tamper-evident model for networks. To evaluate nodes positions according to their impact in a network, we use centralities of nodes. In networks, data stored is not encrypted and therefore more prone to attacks. Attackers compromise a node in a network and due to inter-connectivity of nodes, the attack can result in compromising more nodes in the same network. In some situations, attacker tampers e.g. names or information associated with the individuals or changes the ties/links between the nodes in the network. This thus requires the introduction of mechanisms to revise the structure of the social networks thereby making it more resistant to tampering. Additionally, the mechanism should also  make it evident if such attacks occur. Our proposed method uses a combination of network centrality-based techniques and cryptographic techniques to ensure that the administrator is able to figure out if the network has been tampered with. 
The structure of the rest of the paper is as follows:
First we give background of related areas including an overview of key network centralities. This is then followed by a review of attacks on social networks. Next, we present the results and discussion starting with initialization of the original network and then two cases of modifications - one which was valid and the other being invalid i.e. a case of tampering by a malicious attacker. We then give details of the methods used in the paper giving methodological pipelines and flowcharts of proposed algorithms. Other details of the modeling and analysis are presented in the supplementary information. 

\section*{Background}
Security experts can deploy variety of tools to monitor networks such as Secure Information and Event Management (SIEM) \cite{kotenko2012attack, howell2015building}. Hackers can eavesdrop on network traffic, and tamper with the integrity of information and processes occurring across the network. While on its face, it can appear that changing the nodes and their information may not be easy but in reality, once the attacker has access to the network, the entire network will be compromised. 
Besides, such tampering is not easy to detect at all. The typical mechanism of security involves experts using private keys for encryption employed along with the use of digital signatures, strong authorization, tamper-resistant protocols across communication links. Still, these mechanisms are safe if and only if the keys are not themselves compromised - the processes are not abused. Cryptographic operations in networks might provide a reasonable level of protection for some applications. However, these often involve application on low-level data and thus, can be quite heavy in terms of processing and consuming memory and other valuable resources in big data scenarios. For protecting networks containing sensitive and important information, however, higher levels of assurance are needed. 

\subsection*{Centralities}
Freeman notes that the calculation of centrality is a key area of research focus in the domain of social network analysis for an extended period of time \cite{freeman1979centrality, kimura2009finding}. Most commonly used centrality measures include degree centrality, closeness centrality, betweenness centrality, eccentricity centrality \cite{bouttier2003geodesic} and eigenvector centrality—with degree, closeness and betweenness measures being proposed by Freeman \cite{freeman1979centrality} and eigenvector centrality proposed by Bonacich \cite{bonacich1972factoring}. Centrality is considered important by researchers because centralities formally indicate the value of nodes in the network topology. Central positions have, however, often been equated with opinion leadership or popularity \cite{becker1970sociometric, Roger:2003, valente1996network, valente1999accelerating, askari2013large}. Often, researchers primarily use the degree measure of centrality, perhaps because it is the easiest in terms of explanation to non-technical audiences — besides its association with behavior is intuitive. In the current paper, we are looking to evaluate and validate the role of commonly-used centralities in the identification of nodes which are actually influential in the network.
We focus on the following centralities for the analysis:

\begin{enumerate}
\item Degree Centrality: It is defined as the number of links of node \cite{freeman1979centrality}. Degree centrality of a node v is calculated as: 
\(C_{D}\left(v\right)\doteq\frac{k_{v}}{n-1}=\sum_{j\in G}\frac{a_{vj}}{n-1}\), 
where \(k_{v}\) is the degree of a node, n is the total number of the nodes in the network.

\item Betweenness Centrality: Betweenness centrality quantifies “the number of times a node acts as a bridge along the shortest path between two other nodes” \cite{freeman1979centrality}. Betweenness centrality is calculated as follows:
\(C_{B}(v)=\sum_{s\neq v\neq t}\frac{\sigma_{st}(v)}{\sigma_{st}}\), 
where \(\sigma_{st}\) is total number of shortest paths from node s to node t and \(\sigma_{st}(v) \) is the number of those paths that intersect node v.

\item Closeness Centrality: defines as node closeness towards each node in a network \cite{sabidussi1966centrality}. It is calculated using the formula: 
\(C_{C}(v)=\sum\frac{1}{dist(v,t)}\), where v and t are the nodes from the vertices G.

\item Eccentricity Centrality: The eccentricity centrality of a node is equal to “the largest geodesic distance between the node and any other node” \cite{bouttier2003geodesic}. Generally, when the Eccentricity centrality is higher for a node, the rate of diffusion for the same is lower. It is calculated as follows: 
\(C_{Ecc}(v)=\frac{1}{Max\left\{ \left(dist(v,t)\right)\right\} }\), where v and t are the nodes from the vertices G.

\item Eigenvector Centrality: It is defined as a “Measure of the influence of a node in a network” \cite{bonacich1972factoring}. Eigenvector is defined as follows:
\(\lambda v\doteq Av\), where A is the adjacency matrix of the graph, \(\lambda \) is a constant (the eigenvalue), and v is the eigenvector.
\end{enumerate}

\subsubsection*{A Short Review of Attacks on Social Networks}
With the recent rise in incidents at the global levels,  there has also been a corresponding considerable increase of interest in  research on the spread and tracking of terrorism. Reid and Chen \cite{reid2007mapping}, present an intellectual structure of research conducted from the last two decades. The authors focus on terrorism outbreaks around the world. They present the structure of the research in the domain. Visualization techniques have been used to map contemporary terrorism research domains which includes data mining, analysis, charting, and visualizing the terrorism research area according to experts, institutions, topics, publications, and social networks. Domain mapping is an important but difficult task as it is not easily accessible but it is of clandestine nature. Neither the intellectual structure nor the characteristics are easily traceable. Domain mapping helps in investigating trends and validating perceptions for experts whereas for newbies this mapping provides new research areas as well as being helpful in research development for new areas. The investigations of this proposes that prior researches have a heavy influence on new researches and also mostly cited work such as Rohan Gunaratna and many others are heavily influenced by the previous performed research on terrorism.

Cao \cite{cao2010depth} highlights the importance of behavior analysis in securing systems. These systems could range from those employed for business intelligence to social computing. This is besides usage for the analysis of intrusion detection in networks to events and even in making decisions. Additionally, these could be useful for business analysis as these methods can be used to analyze markets, users preferences and for the detection of exceptional behavior of terrorists and criminals in networks. Traditional ways are used to detect such behaviors but they are not well organized nor the transactional data covers all aspects of representations of human behaviors. The authors have introduced advanced ways of behavior analysis due to inefficiency of traditional analysis. Their methods involve the use of important information such as links between entities which helps in extracting the hidden elements in transactional data. The goal of behavior analysis us to help develop methodologies, techniques, and practical tools for representing, modeling, analyzing and understanding of networks for detecting anomalies. In behavioral network, intrinsic mechanisms change a network from inside which effect in network topological change. One of the major research issues in this domain is that behavioral elements are often intersprersed in transactional data and it can be considerably difficult to gather and analyze them in their entirety. 

Zhu et al. \cite{zhu2010visualizing} note that social network concepts cannot be estimated easily but they can only be visualized through computer simulations. Although there are various tools and techniques available for visualization but they are often unable to perform complete visualization of social network concepts. This paper proposes a new concept based on visualization which them presented in the form of a , "NetVizer" which gives a better visualization of betweenness centrality concepts . Social network analysis is often employed by organizations for data mining, and also for understanding decision-making processes. The idea is to maximize information flow in employee social networks. Social network analysis is being used in various fields such as law firms. medicine companies, and financial institutions besides being used in research and development organizations. The uses of social network analysis based decision making can range from expert assessment, criminal investigation, and community understanding. The analysis is typically focused on the network information.

Besides NetVizer, there are other tools such as proposed by Chung et al. in \cite{chung2005evaluating}. The paper presents a crime analysis tool for dynamic visualization of events. This tools has been proposed as an improvement on previous visualization tools which were manual and less efficient. This tools finds spatial temporal patterns of crimes and visualizes them. Previously, network charts were used for crime analysis and were drawn manually where as other software applications were too difficult to use and interpret. Therefore, there is a need to develop automatic crime detection and analysis tools which are easy to use and interpret data easily.

Other than completely relying on tools, there has to be ways which help in detecting networks loopholes. These include work by Van der Aalst and Medeiros  \cite{van2005process} which suggests the frequent checking of audit trails in any organization to help detect security breaches and anomalies. Audit trails are used for analyzing security violations in systems where processes log their events through time stamps indicating the causality of events by stamping the time of occurrence. The paper presents an alpha-algorithm which can be used to support security at various levels of a systems such as from process execution to checking conformance. This algorithm uses process mining techniques for storing and monitoring audit trails in organizations. 

Chau and Xu  \cite{chau2007mining} present a semi automated approach to identify certain groups by studying their important structural characteristics. People with certain opinions and emotions are studied such as those forming hate groups. Over the internet, there are various social networks which can be used to propagate opinions,  emotions, and beliefs thereby influencing other people. These methods have traditionally gained considerable success. A number of techniques were used to study such networks including web-mining and social network analysis to study crimes over the internet such as in the formation of extremists groups and other terrorist organizations \cite{o2006using,sommer2010outside,oatley2006decision}. 
Social network analysis has this clearly been extremely helpful in exploring networks and their characteristics, organizational and inter-organizational behavior and in many other domains. But also helps in identifying central nodes based on their functioning roles in networks. 

Beside tools and algorithms, models such as those based on clustering  \cite{nath2006crime} can also be used in tracing crime patterns. 

Ahmad et al. \cite{ahmad2009mining} identify gold farmers in gaming social networks. These  are involved in the illegal practice of buying and selling of virtual goods in online games for real money. This employs mining techniques for the detection of such group in networks.

Ball notes that Social Network analysis can be a very effective automated tool in counter-terrorism research \cite{ball2016automating}.  Kukkala et al. demonstrate privacy-preserving social network analysis in distributed social networks  \cite{kukkala2016privacy}. Ongkowijoyo and Doloi have employed social network analysis to understand risks to the infrastructure \cite{ongkowijoyo2017determining}. Colladon and Remondi have used Social Network Analysis to detect and prevent money laundering \cite{colladon2017using}.

\section*{Results \& Discussion}
In this section, we present proof of concept for the proposed model performed on an empirical data set of Zachary Karate Club Network \cite{zachary1977information}. This is followed by a detailed discussion and analysis of various possible scenarios resulting from tampering in the light of the proposed tamper-evident algorithm. 
\subsubsection*{Data sets}
The empirical data set has been collected form a real world network published in \cite{zachary1977information}. This data set contains 34 individuals bonded with each other for forming a social network responsible for diffusing information in the network. The published data set has been used for experimentation so as to provide a proof of concept to the proposed model and algorithms. 

We employ a combination of centralities on the collected network. It is because the centrality measures are used to identify the critical positions of nodes in a network and centralities give mathematical value to these positions. If a position of any node changes in the network either it is the addition or deletion of a vertices or edge, so does the centrality value. We take Hash values of the centrality measures to maintain the integrity of the measured values. 

\subsection*{Building A Tampering Evident Model}

\subsubsection*{Scenario I: Original Network}
\begin{figure}[ht]
\centering

\includegraphics[width=0.7\linewidth]{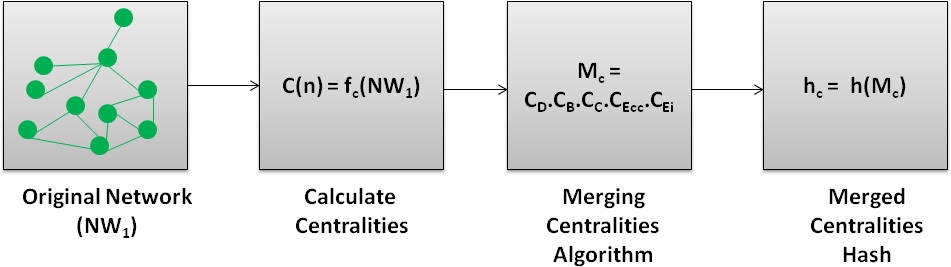}
\caption{Methodology Framework for Original Network.}
\label{fig:Methodology-Framework-to-original}
\end{figure}

We test the original network without altering the integrity of any node and compare the hash value calculated before and after. Figure \ref{fig:Methodology-Framework-to-original} shows the procedure followed to authenticate and protect the integrity of the network. This shows data is used to form a network which is going to be a baseline for rest of the scenarios after which centralities are calculated, \(C(n)\). Then, we use merging algorithm to merge all centralities as even a slightest change in network will effect the centralities values. For preserving the integrity of the network, we calculate hashed values, \(h_{c}\) of the merged centralities, \(M_{c}\). 

\subsubsection*{Scenario II: Valid Modification Network}
We assume here that ties/links between the nodes can be changed over time. This change is authorized and needs to be handled carefully as to differentiate the changes done by attackers. Therefore all the authorized changes are updated regularly by using an update algorithm at regular intervals.

\begin{figure}[ht]
\centering
\includegraphics[width=0.6\linewidth]{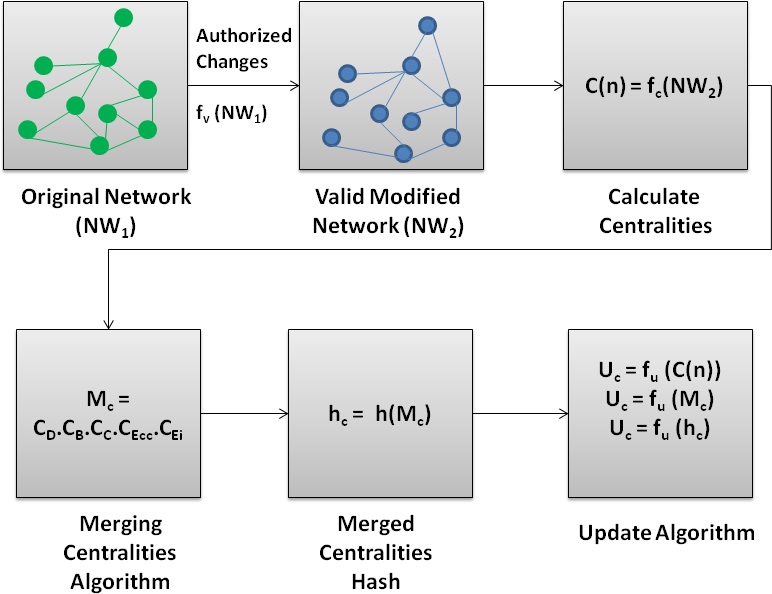}
\caption{Methodology Framework to Valid Modified Network.}
\label{fig:Methodology-Framework-to-valid}
\end{figure}

Figure \ref{fig:Methodology-Framework-to-valid} shows the procedure followed to authenticate and protect the integrity of the network. The network \((NW_{1})\) is used on which authorized changes are performed to form a new network \((NW_{2})\) after which centralities are calculated, C(n). Then, we use merging algorithm to merge all centralities of the new network. For protecting the integrity of the network, we calculate hashed values, \(h_{c}\) of the merged centralities, \(M_{c}\).We use update algorithm to store new calculated values of the network \((NW_{2})\). We added and deleted links naturally and updated the centrality values simultaneously and verified the hash values to see if the suggested model can work for dynamic networks.

\subsubsection*{Scenario III: Tampered Network}
For this scenario, we have two situations. One is attacker compromises a single node in the network where as second scenario is to avoid detection at any point, he deletes the whole network.
For this scenario, we have two situations. One is attacker compromises a single node in the network where as second scenario is to avoid detection at any point, he deletes the whole network.

\begin{figure}[ht]
\centering
\includegraphics[width=0.7\linewidth]{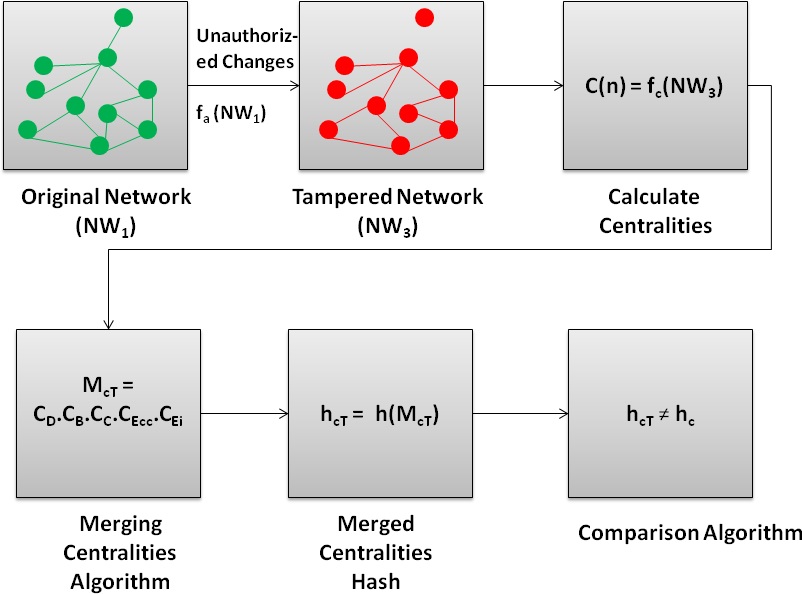}
\caption{Methodology Framework to Tampered Network.}
\label{fig:Methodology-Framework-to-Tamper}
\end{figure}

In figure \ref{fig:Methodology-Framework-to-Tamper} shows formal procedure followed to authenticate and protect the integrity of the network. The network \( (NW_{1})\)  is used on which attacker changes are performed which forms a new network \((NW_{3})\) after which centralities are calculated, \(C(n)\)  for the attacked network. Textual merge algorithm is used to merge all centralities of the attacked network. For evaluating the integrity of the network, we calculate hash values, \(h_{cT}\) of the textual merged centralities of attacked network, \(M_{cT}\). We use comparison algorithm to compare the new calculated hash values of attacked network \((NW_{3})\) with the stored values of original network i.e \(h_{c\text{ }}\text{\ensuremath{\neq}}h_{cT}\).

Our experiments demonstrate that each of the centrality measures has a unique effect nodes of networks. Therefore, our proposed model can easily detect minor changes in the network structure.

\subsection*{Features of Tamper-evident Model}
We focus to propose tamper-evident model driven by three main objectives:
\begin{enumerate}
\item Tamper detection: Any unauthorized changes will be detected and will also indicate the source of change occurred.
\item Independence: The proposed model does not require any cooperation from monitored systems and does not depend on other components installed for monitoring network activities.
\item Lightweight verification: The proposed model is efficient for monitoring network and and can be easily integrated into systems and other system applications.
\end{enumerate}
 
 \subparagraph*{Theorem 1}

Tamper-evident networks algorithms can be implemented with a time complexity of \(O(n)\).

Our proposed model satisfies the mentioned goals. For this, we have applied the proposed mechanism on empirical networks. Details are given in the section on "Methods".

\section*{Methods}
In this paper, we propose cognitive digital foot-printing in social network in building a tamper-evident model. For this we have applied the basic digital foot printing concept in which any unauthorized tampering done to any node will be easily detectable. 

In this section we present the tamper-evident storage mechanism in the form of several algorithms. 

\clearpage
\subsection*{Main Algorithm}
Following presents the Main algorithm which defines the overall working of the proposed Tamper-evident model. This involves further algorithms explained later.

The proposed algorithm functionality can be described as following:
\begin{enumerate}
\item Generate the network from collected data for network formation.
\item We run Node-Safe Hash (NSH) algorithm, discussed later.
\item Next, system will check if the changes are done by an authenticated/valid user or not. 
\item If the changes are from a valid user, system will execute Tamper-check (TC) algorithm, discussed later.
\item TC algorithm checks for tampering in the network. If the network has been tampered, this will generate an alarm and notify the admin to take an appropriate action. Else, this updates the new hash values to old hash values and the loop continues.
\end{enumerate}

\begin{figure}[!ht]
  \centering

\includegraphics[width=0.3\linewidth]{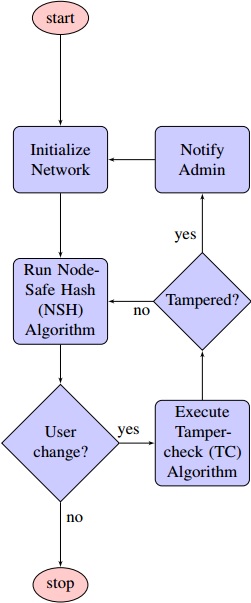}
     \caption{Main Algorithm.}
     \label{fig:Main-Algorithm}
     \end{figure}
     
\clearpage
\subsection*{Node-Safe Hash (NSH) Algorithm}
Node-safe hash algorithm is used to calculate hash value of nodes and saving it appropriately. This follows as:
\begin{enumerate}
\item Take the complete data of nodes collected for network formation.
\item Execute the Calculate Centralities (CC) algorithm, discussed later.
\item Take all nodes and save their calculated hash values.
\item Check if it is not a last node then take the next to save its associated hash, else, stop.
\end{enumerate}


\begin{figure}[!ht]
  \centering

\includegraphics[width=0.3\linewidth]{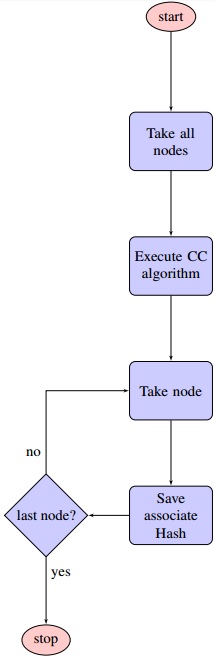}
   
     \caption{Node-Safe Hash (NSH) Algorithm.}
     \label{fig:Node-Safe-Hash-(NSH)}
     \end{figure}

\clearpage
\subsection*{Calculate Centralities (CC) Algorithm} 

Calculate Centralities algorithm calculates all centralities hash values by the following proposed procedure:
\begin{enumerate}
\item Calculate all the centralities which includes Degree Centrality \(C_{D}\) , Betweenness Centrality \(C_{B}\), Closeness Centrality \(C_{C}\), Eccentricity Centrality \(C_{Ecc}\), Eigenvector Centrality \(C_{Ei}\).

\item We perform textual merge of centralities to combine the centralities.

\item Then, apply SHA1 algorithm to the calculated centralities for keeping the integrity-check of merged centralities.

\end{enumerate}


\begin{figure}[!ht]
  \centering
 \includegraphics[width=0.8\linewidth]{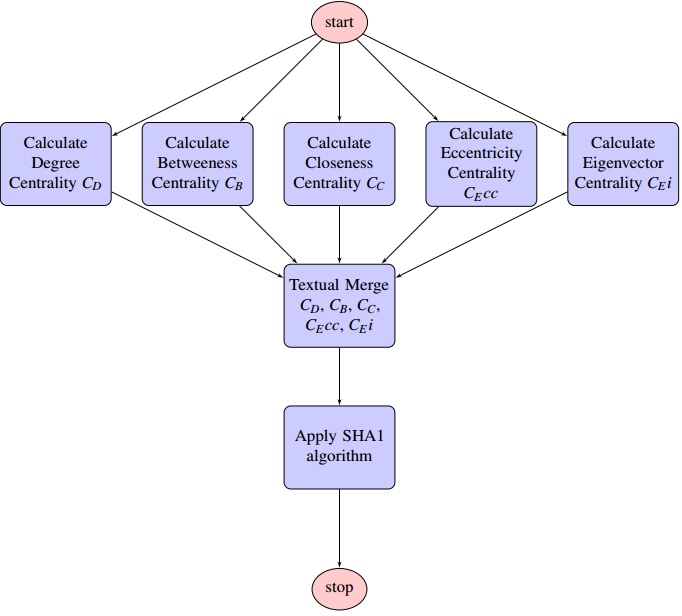}        
   \caption{Calculate Centrality (CC) Algorithm.}
     \label{fig:Calculate-Centralities-(CC)}
     \end{figure}

\clearpage
\subsection*{Tamper-Check (TC) Algorithm}

Following describes the working of Tamper-check algorithm. This checks if the network has been tampered or not. This works as following: 
\begin{enumerate}
\item We run NSH algorithm to calculate and save the hash values of network nodes.
\item Check if the calculated hash value is equal to the previously calculated hash values. 
\item If the new calculated hash values are not equal to old hash values then network is being tampered else, it is not. 
\end{enumerate}


\begin{figure}[!ht]
  \centering
\includegraphics[width=0.3\linewidth]{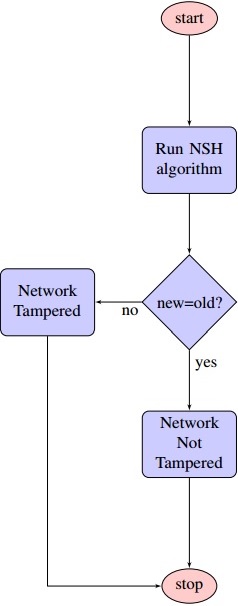}
   \caption{Tamper-check (TC) Algorithm.}
     \label{fig:Tamper-check-(TC)-Algorithm}
     \end{figure}
\clearpage
\subsection*{Calculate SHA1 Algorithm}

SHA1 is used for preserving the integrity of data contents \cite{eastlake2001us}. We use hashing algorithms to authenticate if the contents have been changed or modified to detect masqueraders who insert message from fraudulent sources. This also detects if the content is modified by insertion, deletion and reordering sequence and also notifies is timing modification is done which is used for replaying valid sessions. This algorithm follows as in \cite{eastlake2001us} and has been implemented in our proposed solution as:

\begin{enumerate}
\item Take the textual merge of centralities as input. 
\item Add padding bits to input value to make it congruent to 448 mod 512. That is adding one 1 and as many 0's to make it congruent to 448 mod 512.
\item Append a 64 bits length to the the padded textual merged centralities. These bits hold the binary format of 64 bits indicating the length of the original message. 

\item Prepare Processing functions - This requires 80 processing functions defined as following \cite{eastlake2001us}: 

\(f(t;B,C,D) = (B AND C) OR ((NOT B) AND D) ( 0 <= t <= 19)\)
\(f(t;B,C,D) = B XOR C XOR D (20 <= t <= 39)\)
\(f(t;B,C,D) = (B AND C) OR (B AND D) OR (C AND D) (40 <= t <=59)\)
\(f(t;B,C,D) = B XOR C XOR D (60 <= t <= 79)\)

\item Prepare Processing Constants - 80 processing constants are required to produce 5 words, defined as following \cite{eastlake2001us}:

\(K(t) = 0x5A827999 ( 0 <= t <= 19)\)
\(K(t) = 0x6ED9EBA1 (20 <= t <= 39)\)
\(K(t) = 0x8F1BBCDC (40 <= t <= 59)\)
\(K(t) = 0xCA62C1D6 (60 <= t <= 79)\)

\item Initialize Message Digest Buffers - In this integrity check algorithm, SHA1 requires 160 bits or 5 buffers of words (32 bits) \cite{eastlake2001us}: 

\(H0 = 0x67452301\)
\(H1 = 0xEFCDAB89\)
\(H2 = 0x98BADCFE\) 
\(H3 = 0x10325476\)
\(H4 = 0xC3D2E1F0\)

\item Processing message in 512-bit blocks (L blocks in total message) - This is the core functioning of SHA1 algorithm which loops through the padded and appended message in 512-bit blocks. Input and predefined functions: \(M[1, 2, ..., L]\): Blocks of the padded and appended message \(f(0;B,C,D), f(1,B,C,D), ..., f(79,B,C,D)\): 80 Processing Functions \(K(0), K(1), ..., K(79)\): 80 Processing Constant Words \(H0, H1, H2, H3, H4, H5\): 5 Word buffers with initial values.

\end{enumerate}

SHA1 works as: 

For loop on \(k = 1 to L (W(0),W(1),...,W(15)) = M[k] /* Divide M[k] into 16 words */\)

For t = 16 to 79 
do: \(W(t) = (W(t-3) XOR W(t-8) XOR W(t-14) XOR W(t-16)) <<< 1 A=H0, B=H1, C=H2, D=H3, E=H4\)

For t = 0 to 79 

do: \(TEMP = A<<<5 + f(t;B,C,D) + E + W(t) + K(t) E=D, D=C, C=B<<<30, B=A, A=TEMP\) 

End of for loop

\(H0=H0 + A, H1=H1 + B, H2=H2 + C, H3=H3 + D, H4=H4 + E\)
End of for loop 

Output: \(H0, H1, H2, H3, H4, H5\): Word buffers with final message digest.

\begin{figure}[!ht]
  \centering
 \includegraphics[width=0.8\linewidth]{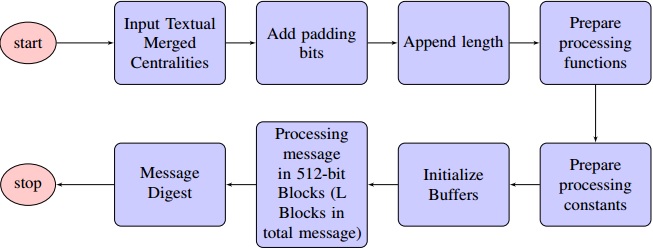}
   \caption{Secure Hash (SHA1) Algorithm.}
     \label{fig:SHA1-Algorithm}
     \end{figure}
\clearpage

\section*{Conclusion}
In this paper, we have presented a tamper evident mechanism for complex networks, in general, and complex genomic networks, in particular. We have demonstrated the model using  a proof-of-concept validation scheme using actual data sets. Our proposed model detects unauthorized changes in a given network. Experiments have also been presented which were  carried out on a real world network to allow for an examination of how network changes can be easily detected using the proposed approach. We have used centralities which are more commonly used for detecting critical positions of nodes in complex networks. The experiments clearly demonstrate that the model can be easily implemented in complex networks for detecting any changes in the network. We have focused here primarily on the network structure and not on the  node attributes or node information. Our experiments results show that even a minor change in network structure can be easily detectable through this model. A limitation of the proposed model is that currently it works only on the network structure and not on the node attributes or information. In the future, the proposed technique can be extended to include node-based information. Cryptographic techniques can be further included in the proposed mechanisms.

\bibliography{TampBib}

\section*{Author contributions statement}

M.N. conceived the experiments and analyzed the results, K.B. conducted the experiments and analyzed the results. All authors reviewed the manuscript. 

\section*{Competing interests}
The authors declare that they have no competing financial interests.

\section*{Correspondence}
Correspondence and requests for materials should be addressed to Dr. Muaz A. Niazi (email: muaz.niazi@ieee.org).


\end{document}